\documentstyle[floats,preprint,aps,epsf,amssymb]{revtex}

\begin{document}

\title{Oscillatory null singularity inside realistic spinning black holes}

\author{Amos Ori}
\address {Department of Physics,
          Technion---Israel Institute of Technology, Haifa, 32000, Israel}
\date{\today}
\maketitle

\begin{abstract}
We calculate the asymptotic behavior of the curvature scalar 
$(Riemann)^2$
 near the null weak singularity at the inner horizon of a generic spinning 
black hole, and show that this scalar oscillates infinite number of times while 
diverging. The dominant parallelly-propagated Riemann components oscillate 
in a similar manner. This oscillatory behavior, which is in a remarkable 
contrast to the monotonic mass-inflation singularity in spherical charged 
black holes, is caused by the dragging of inertial frames due to the black-
hole's spin.

\end{abstract}
One of the major challenges in Classical general relativity during the last few 
decades has been to explore the nature of the spacetime singularities which 
form in gravitational collapse. The existence of singularities inside black 
holes has been verified by several mathematical theorems \cite{1}. 
However, the singularity theorems do not tell us much about the location 
and features of these singularities. 

It is widely anticipated that the realistic astrophysical black holes are 
rapidly spinning \cite{2,3}. The simplest type of a spinning black hole (BH) is 
given by the Kerr solution \cite{4}, describing a stationary, axially-
symmetric, spinning vacuum BH. The inner horizon (IH) is a null hypersurface 
located 
inside the BH. This hypersurface, also known as the Cauchy horizon \cite{1}, 
marks the boundary of predictability for physical fields whose initial data 
are specified outside the BH. In the pure Kerr geometry, the IH is a perfectly 
smooth surface. Penrose \cite{5} pointed out, however, that ingoing 
electromagnetic or gravitational perturbations are infinitely blue-shifted at 
the IH. He therefore suggested that in a more realistic BH, which is not 
strictly stationary, the infinitely blue-shifted perturbations will lead to the 
formation of a curvature singularity instead of a regular IH (we shall refer 
to this singularity as the IH singularity). The instability of the inner horizon 
was later investigated by several authors, who used a spherical charged BH 
as a toy model \cite{6} (this is a useful toy model, because a spherical 
charged BH also admits an inner horizon with infinite blue shift). A few 
analyses of linear fields inside a Kerr BH have also been carried out at the 
end of the 1970's \cite{7,8}. (For recent analyses of the late-time behavior 
of gravitational perturbations outside a Kerr BH, see \cite{9}.)

About ten years ago, in an effort to explore the non-linear aspects of 
the IH singularity, Poisson and Israel \cite{6} introduced the mass-inflation 
model -- a spherically-symmetric model made of a charged BH with two 
radial null fluids (ingoing and outgoing). In this model they obtained a null 
curvature singularity at the IH, known as the mass-inflation singularity. This 
singularity is marked by an exponential growth of curvature. A more detailed 
study \cite{10} later revealed that the mass-inflation singularity is weak in 
Tipler's \cite{11} terminology. Namely, physical objects only experience 
finite tidal distortion when they approach the singularity. 

Later, Ori \cite{12} investigated the geometry inside a realistic 
spinning BH using a perturbative approach (see also \cite{13}). This analysis 
revealed that in the spinning case, too, there is a null, weak, curvature 
singularity at the IH. The main results of the perturbative analysis \cite{12} 
were later confirmed by several non-perturbative local analyses 
\cite{14,15,16,17}. 

In general, the features of the IH singularity of spinning BHs are found 
to be very similar to that of spherical charged BHs: In both cases, the 
singularity is null, weak, and blue-shift dominated. There is one important 
difference, however: The mass-inflation singularity is characterized by a 
monotonic growth of the mass function (and curvature) \cite{6,10}. On the 
contrary, the IH singularity of a spinning BH is oscillatory, as we shall show 
in this paper. This oscillatory behavior is related to the dragging of inertial 
frames due to the BH's spin.

One of the rather surprising findings of the perturbation analysis 
\cite{12} is that the IH-singularity is essentially linear. Namely, at the early 
portion of the IH, the structure of the singularity may adequately be 
described (at the leading order) by the linear gravitational perturbation over 
the Kerr background, because the effect of higher-order non-linear 
perturbation terms is negligible. Motivated by this observation, we have 
recently carried out a detailed analysis \cite{18} of linear gravitational 
perturbations over the Kerr background, using the Newman-Penrose (NP) 
formalism. Based on the results of this analysis (along with that of Ref. 
\cite{12}), we shall now calculate the asymptotic behavior of the curvature 
at the IH-singularity and reveal its oscillatory character. For concreteness, 
we shall focus on the quadratic curvature scalar $K\equiv R_{\alpha \beta 
\gamma \delta }R^{\alpha \beta \gamma \delta }$
. We shall consider a non-extreme, pure vacuum BH, and restrict attention 
to the early portion of the IH singularity (where the perturbation analysis 
\cite{12} is effective).

The event horizon and the IH of the background Kerr geometry are 
located at the hypersurfaces 
$r=r_+$
 and 
$r=r_-$, respectively, where $r_\pm \equiv M\pm (M^2-a^2)^{1/ 2}$
. Here $M$ and $a$ denote the BH's mass and specific angular momentum, 
respectively. We use here the Boyer-Lindquist \cite{19} coordinates 
$(t,r,\theta ,\varphi )$
. The Eddington-like coordinates $u,v$ are given by $v=r*+t$ and $u=r*-t$, 
where $r*(r)$ is defined by $dr/ dr*=\Delta / (r^2+a^2)$
 and 
$\Delta \equiv (r-r_+)(r-r_-)$
. The event horizon and the IH correspond to $u=-\infty $
 and $v=\infty $
, respectively. 

Following Ref. \cite{12}, we express the metric 
$g_{\alpha \beta }$
 of the perturbed spinning BH as the sum of the unperturbed Kerr metric 
and the metric perturbation 
$h_{\alpha \beta }$
. The latter is then expanded in the form
\begin{equation}
h_{\alpha \beta }=h_{\alpha \beta }^{(1)}+h_{\alpha \beta 
}^{(2)}+h_{\alpha \beta }^{(3)}+...
\, 
\label{1}
\end{equation}
where 
$h_{\alpha \beta }^{(1)}$
 is the linear metric perturbation, 
$h_{\alpha \beta }^{(2)}$
 is the second-order perturbation, etc.. We adopt here the gauge used in 
Ref. \cite{12}, in which all terms 
$h_{\alpha \beta }^{(J)}$
 are finite at the IH (and are arbitrarily small at its early portion), and the 
null curvature singularity is located strictly at 
$r=r_-$
 (i.e. at the IH) of the Kerr background. This singularity is marked by the 
divergence of the curvature scalar K. Note that K (like the Riemann tensor 
itself) is perfectly regular at the IH of the unperturbed Kerr background, 
and its divergence in a realistic spinning BH is caused by the gravitational 
perturbation, which is infinitely blue-shifted at the IH. 

Comparing the asymptotic forms of the various terms 
$h_{\alpha \beta }^{(J)}$
, one finds that 
$h_{\alpha \beta }$
 is dominated by the linear perturbation 
$h_{\alpha \beta }^{(1)}$
 \cite{12}; The higher-order terms are smaller by certain powers of $1/v$ 
and/or $1/u$ (which is arbitrarily small at the early portion of the IH.) As a 
consequence, it is not difficult to show that K is dominated by 
$\hat K\equiv \hat R_{\alpha \beta \gamma \delta }\hat R^{\alpha \beta 
\gamma \delta }$,
where 
$\hat R_{\alpha \beta \gamma \delta }$
 denotes the linear perturbation in the Riemann tensor. In what follows we 
shall use the NP formalism to calculate 
$\hat K$
.

In a vacuum spacetime, the Riemann tensor may be expressed as a 
linear combination of the five NP Weyl scalars 
$\Psi _i$
 (i=0...4) and their complex conjugate (see e.g. Eq. (1.298) in \cite{20}). We 
schematically write this linear combination as 
\begin{equation}
R_{\alpha \beta \gamma \delta }=Q_{\alpha \beta \gamma \delta }^i\Psi 
_{\kern 1pt i}+c.c.
\ ,
\label{2}
\end{equation}
where 
$Q_{\alpha \beta \gamma \delta }^i$
 are constants, and c.c. denotes the complex conjugate. Explicitly calculating 
these constants according to the method explained in Ref. \cite{20}, and 
then squaring the last equation, one finds
\begin{equation}
R_{\alpha \beta \gamma \delta }R^{\alpha \beta \gamma \delta }=8\left( 
{\Psi _0\Psi _4+3\Psi _2^2-4\Psi _1\Psi _3} \right)\,+c.c.
\ .
\label{3}
\end{equation}
In a similar manner, by picking the linear perturbations of the quantities in 
both sides of Eq. (\ref{2}) and squaring them, one obtains an analogous 
expression for 
$\hat R_{\alpha \beta \gamma \delta }$
:
\begin{equation}
\hat K\equiv \hat R_{\alpha \beta \gamma \delta }\hat R^{\alpha \beta 
\gamma \delta }\cong 8\,\left( {\psi _0\kern 1pt \psi _4+3\psi _2^2-
4\psi _1\psi _3} \right)\,+c.c.
\ ,
\label{4}
\end{equation}
where 
$\psi _i$
 denotes the linear perturbation in 
$\Psi _i$
. [We have ignored here all contributions proportional to the 
(undifferentiated) metric perturbation 
$h_{\alpha \beta }$
, e.g. those obtained when indices are raised or lowered. These turn out to 
be negligibly small, like 
$h_{\alpha \beta }$
 itself.] 

From the asymptotic expressions for the linear metric perturbations 
\cite{12}, one can evaluate the maximal possible divergence rates of the 
various linear NP fields at the IH. One can show that the maximal inverse 
powers of 
$r-r_-$
 involved in this divergence are \cite{21}
\begin{equation}
\psi _0\propto (r-r_-)^{-2}\quad ,\quad \psi _1\propto (r-r_-)^{-1}\quad 
,\quad \psi _{2,3,4}\propto (r-r_-)^0
\ .
\label{5}
\end{equation}
Therefore, 
\footnote{In the gauge we use, 
$\Psi _2$
 is dominated by its second-order term, 
$\Psi _2^{(2)}$
, which diverges like 
$(r-r_-)^{-1}$
 [whereas 
$\Psi _2^{(1)}\equiv \psi _2\propto (r-r_-)^0$
]. Nevertheless, the contribution of 
$\Psi _2^{(2)}$
 (squared) to K is smaller than $\hat K$
 by certain powers of $1/u$  or $1/v$, as was mentioned above (with 
regards of the contribution of non-linear perturbations to K). All other NP 
fields 
$\Psi _i$ ($i\ne 2$
) are dominated by their linear counterparts, 
$\psi _i$}
\begin{equation}
K\cong \hat K\cong 8\,\psi _0\,\psi _4+c.c.
\ .
\label{6}
\end{equation}
This result is remarkable for two reasons. First, 
$\psi _0$ and $\psi _4$
 are gauge-invariant \cite{20} (whereas 
$\psi _{1,2,3}$
 are not; the expressions for 
$\psi _{1,2,3}$
 in Eq. (\ref{5}) are obtained in the gauge used in Ref. \cite{12} and here.) 
Second, both 
$\psi _0$ and $\psi _4$
 satisfy a simple master equation \cite{22}.

The evolution of 
$\psi _0$ and $\psi _4$
 inside a Kerr BH was analyzed in Ref. \cite{18} 
\footnote{Note that 
$\Psi ^{s=2}$
 and 
$\Psi ^{s=-2}$
 therein correspond respectively to 
$\psi _0$ 
 and 
$(r-ia\,\cos \theta )^4\,\psi _4$
 in the notation of the present paper.}
 . For generic initial data, one finds that both 
$\psi _0$ and $\psi _4$
 are dominated by the modes with l=2 (which have the slowest decay rate, 
$t^{-7}$
, outside the BH). The asymptotic behavior of 
$\psi _4$
 at the early portion of the IH is found to be (see section IX in Ref. 
\cite{18})
\begin{equation}
\psi _4\cong u^{-8}\,(r_--ia\,\cos \theta )^{-4}\sum\limits_{m=-2}^2 
{\,A_m\,_{-2}Y_2^m(\theta ,\phi )\,e^{-im\Omega _-u}}+O\,(u^{-9})
\ ,
\label{7}
\end{equation}
where 
$\Omega _-\equiv a/ (2Mr_-)$
, $\phi \equiv \varphi -\Omega _-t$
 is an azimuthal coordinate regular at the IH \cite{20}, and 
$_sY_l^m$
 denotes the spin-weighted spherical harmonics. The asymptotic behavior of 
$\psi _0$ 
 is
\begin{equation}
\psi _0\cong (r-r_-)^{-2}\,v^{-7}\sum\limits_{m=-2}^2 
{\,B_m\,_2Y_2^m(\theta ,\phi )\,e^{im\Omega _-v}}+O(v^{-8})
\ .
\label{8}
\end{equation}
$A_m$
 and 
$B_m$
 are coefficients that are proportional to the initial amplitudes of the modes 
(l=2, 
$|m|\,\le 2$
) of 
$\psi _4$
 and 
$\psi _0$ 
, respectively. These coefficients are generically non-vanishing. The only 
exception is the coefficient 
$B_0$
, which vanishes identically (that is, the mode l=2,m=0 of 
$\psi _0$ 
 is 
$\propto (r-r_-)^{-2}v^{-8}$
 at the IH; see Ref. \cite{18}). Substituting Eqs. (\ref{7}) and (\ref{8}) into 
Eq. (\ref{6}), we obtain
\begin{equation}
K\cong (r-r_-)^{-2}\,v^{-7}\,\sum\limits_{m=1,2} {C_m}(u,\theta ,\phi 
)\,e^{im\Omega _-v}\,+c.c.
\ ,
\label{9}
\end{equation}
with (generically) non-vanishing coefficients 
$C_m$
. Note that no m=0 term is present at the leading order, due to the 
vanishing of $B_0$
.

Consider now a freely-falling observer which hits the IH singularity at 
a point 
$(u_0,\theta _0,\phi _0)$
. For this observer, 
$r-r_-$
 and $v$ are proportional to
$\tau $
 and 
$\ln \,\left( {-\tau / M} \right)$
, respectively, where 
$\tau $
 denotes the proper time, and we have set 
$\tau =0$
 at the intersection with the IH singularity. One obtains
\begin{equation}
K\cong c\,\tau ^{-2}\left[ {\ln \,\left( {-\tau / M} \right)} \right]^{\,-
7}\,\sum\limits_{m=1,2} {C_m}(u_0,\theta _0,\phi _0)\,\,e^{-imp\ln \left( 
{-\tau / M} \right)}\,+c.c.
\ ,
\label{10}
\end{equation}
where c is a non-vanishing constant that depends on the geodesic's 
constants of motion, and 
$p\equiv a\,(M^2-a^2)^{-1/ 2}$
. In a similar manner, one finds that the most divergent components of the 
Riemann tensor (as measured by a parallelly-propagated tetrad) are 
$\propto \psi _0$
, and are hence proportional to
\begin{equation}
\tau ^{-2}\,\left[ {\ln \left( {-\tau / M} \right)} \right]^{-
7}\,\sum\limits_{m=1,2} {c_m\,_2Y_2^m}(\theta _0,\phi _0)\,\,e^{-
imp\ln \left( {-\tau / M} \right)}\,+c.c.
\label{11}
\end{equation}
with non-vanishing constants 
$c_m$
.

From Eq. (\ref{10}) it is obvious that, while diverging like $\tau ^{-2}$
 (softened by an inverse-power logarithmic factor), the curvature scalar K 
undergoes infinite number of oscillations. In particular, K vanishes and 
changes sign infinitely many times on the approach to the IH-singularity. The 
dominant parallelly-propagated Riemann components, given in Eq. (\ref{11}), 
behave in a similar manner. Thus, the IH-singularity of a generic spinning BH 
is oscillatory. 

This oscillatory behavior is in a remarkable contrast to the monotonic 
increase of the mass-function (and curvature) in the mass-inflation 
singularity of spherical charged BHs. The oscillations are caused by the 
dragging of inertial frames, due to the BH's angular momentum. More 
specifically, the dragging of the nonaxially-symmetric modes (which 
dominate 
$\psi _0$ 
) leads to oscillations in $v$.

It has been argued by Belinsky, Khalatnikov, and Lifshitz (BKL) 
\cite{23} that a generic singularity (the BKL singularity) exists in the 
solutions of the vacuum Einstein equations which is spacelike and oscillatory. 
Recent numerical and analytical investigations provide further evidence for 
the existence of such singular vacuum solutions \cite{24}. It is remarkable 
that both known generic singularities -- the BKL singularity and the spinning 
inner-horizon singularity -- are oscillatory. Note, however, that apart from 
this common non-monotonic character, these two singularities are very 
different from each other: The BKL singularity is spacelike, strong, and 
extremely complicated (perhaps even chaotic), whereas the inner-horizon 
singularity is null, weak, and of a rather simple asymptotic form.

There also is an important difference in the status of these two 
singularities in connection with their actual occurrence in realistic 
gravitational collapse (or, at least, in connection with our present knowledge 
about their actual occurrence). The actual formation of the null weak inner-
horizon singularity in a generic gravitational collapse has been verified in an 
explicit manner by the perturbative analyses \cite{12,13,18}. (The local 
consistency and genericity of this singularity have been verified also by 
several non-perturbative local analyses \cite{14,15,16,17}.)  On the other 
hand, the analyses of the BKL singularity indicated the local consistency of 
this singularity, and probably also its inevitable occurrence in certain 
cosmological models, but so far not in asymptotically-flat situations. There 
certainly exist generic asymptotically-flat initial-data sets which do not 
develope a BKL singularity (e.g. any set with a sufficiently weak initial field, 
such that no black hole forms). One may attempt to conjecture that 
generically any asymptotically-flat initial-data set which develops a black 
hole will also develope a BKL singularity inside it, but we are not aware of any 
compelling evidence for such a conjecture (recall also that the predictive 
power of the singularity theorems is exhausted by the null inner-horizon 
singularity, which definitely forms in a generic collapse). In fact, for reasons 
which are beyond the scope of this paper, the present author believes that 
the above conjecture is incorrect (but a weaker version, which puts 
restrictions on the spatial topology, may be attempted).

Recent numerical studies of spherical charged BHs perturbed by a 
self-gravitating scalar field indicate that a generic spacelike singularity 
forms when the area of the mass-inflation singularity shrinks to zero 
\cite{25}. It is unclear, however, whether an analogous spacelike singularity 
will form in realistic spinning black holes, which are non-spherical, and which 
have no scalar field. Note that the above scalar-field spacelike singularity is 
monotonic \cite{26}. This type of generic spacelike singularity probably has 
no counterpart in vacuum spacetimes: Both the original work by BKL 
\cite{23} and the recent analyses by Berger and collaborators \cite{24} 
suggest that there exists no generic, monotonic, spacelike, vacuum 
singularity. Therefore, although there is likelihood that a BKL-like spacelike 
singularity will form inside realistic spinning black holes, this is still far from 
obvious, and further research is required in order to clarify this issue.

This research was supported in part by the United States-Israel Binational 
Science Foundation, and the Fund for the Promotion of Research at the 
Technion.


\begin{thebibliography}{99}
\bibitem{21} Equation (\ref{5}) does not include the finer details like e.g. 
the inverse powers of $u$ and $v$; These details are presented below for 
the fields 
$\psi _0$ and $\psi _4$
 which dominate K. Note also that Eq. (\ref{5}) should in principle be 
regarded as an upper bound only. However, as we show below (based on the 
detailed analysis of NP perturbations \cite{18}), for the relevant fields 
$\psi _0$ and $\psi _4$
 these inverse powers are indeed realized.
\end{thebibliography}
\end{document}